\documentclass[aps,prb,twocolumn,floatfix,showpacs]{revtex4}
\usepackage{amsmath,amssymb,graphicx,bm}
\begin{document}
\title{Mean-field theories for disordered electrons: Diffusion pole
  and Anderson localization}

\author{V.  Jani\v{s}} \author{ J. Koloren\v{c}}

\affiliation{Institute of Physics, Academy of Sciences of the Czech
  Republic, Na Slovance 2, CZ-18221 Praha 8, Czech Republic}
\email{janis@fzu.cz, kolorenc@fzu.cz}

\date{\today}


\begin{abstract}
We discuss conditions to be put on mean-field-like theories to be   able
to describe fundamental physical phenomena in disordered   electron
systems. In particular, we investigate options for a   consistent
mean-field theory of electron localization and for a   reliable
description of transport properties. We argue that a   mean-field theory
for the Anderson localization transition must be   electron-hole symmetric
and self-consistent at the two-particle   (vertex) level. We show that such
a theory with local equations can   be derived from the asymptotic limit to
high spatial dimensions. The   weight of the diffusion pole, i.~e., the
number of diffusive states at the Fermi energy,   in this mean-field
theory decreases with the increasing disorder   strength and vanishes in
the localized phase.  Consequences of the   disclosed behavior for our
understanding of vanishing of electron   diffusion are discussed.
\end{abstract}

\maketitle 
\section{Introduction}\label{sec:Intro}

Mean-field theories play an important role in the description of
thermodynamic systems. They are intended and used as a first
approximation offering a qualitative picture of the physics of the
studied phenomena. The mean-field concept has developed from its
initial intuitive ideas of van der Waals and Weiss through the Landau
theory of critical phenomena to its present sophistication and
systematics provided by the limit to infinite-dimensional lattice
models. At present, a modern mean-field theory is no longer a
weak-coupling approximate treatment neglecting spatial fluctuations.
It represents a comprehensive theory providing a phase diagram in the
whole range of the input parameters and simulating the exact behavior
in specific limiting situations.  Without a mean-field theory we are
mostly unable to identify the relevant fluctuations the mean values of
which are reflected by thermodynamic (order) parameters.  Mean-field
theory is particularly important for critical phenomena with divergent
correlation functions, where it allows us to handle singularities in a
consistent and manageable way and to select the proper low-temperature
phase, at least above the lower critical dimension.

Mean-field theories were primarily developed for collective phenomena
in  interacting systems. Nontrivial and sometimes not easily
understandable effects are, however, also induced by randomness.
Randomness, in connection with interaction or with quantum interference,
can cause significant and sometimes even unexpected changes in the
behavior of the system. Since mostly no exact solutions are available for
disordered systems, a mean-field approximation has become one of the most
powerful tools to handle fluctuations in the chemical composition of
solids.
 
Milestones of a mean-field theory for disordered (noninteracting)
electron systems were laid at the end of the sixties and the beginning
of the seventies of the last century. The so-called Coherent Potential
Approximation (CPA) developed at that time is a self-consistent
approximation describing rather accurately the electronic structure
and thermodynamic properties of random alloys not only at the model
level but also in realistic settings.\cite{Elliot74} Later on, the CPA
was shown to fit the modern definition of the mean-field theory as an
exact solution of the model system in infinite spatial
dimensions.\cite{Vlaming92,Janis92}
 
The coherent potential approximation is nowadays considered as a
paragon for mean-field theories of quantum disordered and interacting
systems. Its generalized form\cite{Janis89} offers one possible
interpretation of equations of motion in the Dynamical Mean-Field
Theory (DMFT).\cite{Janis91} In spite of the proved reliability of the
CPA to produce an accurate equilibrium electronic structure of
disordered systems, it fails in encountering for inter-site quantum
coherence and backscattering effects. The CPA is essentially unable to
go beyond the semi-classical description of transport properties
qualitatively captured by the Boltzmann equation. This inability is
due to the fact that the CPA does not include vertex corrections to
the electrical conductivity independently of how strong the disorder
may be.\cite{Velicky69} It is hence unsuitable for the description of
one of the most prominent features of disordered systems: Anderson
localization.

Anderson localization in disordered or amorphous solids takes place
when there are available electronic states at the Fermi surface but no
diffusion or charge transport at long distances is observed.
Possibility of the absence of diffusion in impure metals and alloys
was proposed by P. W. Anderson on a simple tight-binding model of
disordered noninteracting electrons.\cite{Anderson58} Since then,
vanishing of diffusion, now called Anderson localization, has
attracted much attention of both theorists and
experimentalists.~\cite{Lee85,Kramer93} In spite of a considerable
portion of amassed experimental data, disclosed various specific and
general aspects of the Anderson metal-insulator transition, and
a~number of theoretical and computational approaches so far developed
we have not yet reached complete understanding of Anderson
localization. Although many features of the critical behavior at the
Anderson localization transition have been disclosed, the position of
this disorder-driven metal-insulator transition within the standard
classification scheme of phase transitions with control and order
parameters has remained unclear. It has been mainly due to the
nonexistence of an appropriate mean-field-like theory for this
phenomenon. Only very recently the present authors demonstrated that a
mean-field-like solution for the Anderson metal-insulator transition
can be derived from the asymptotic limit to high (but finite) spatial
dimensions.\cite{Janis04a} This solution is very close to the CPA in
the resulting electronic structure, it reduces to the CPA in infinite
dimensions. It, however, differs from the CPA significantly in
transport properties derived from two-particle functions.  In addition
to the one-electron self-consistence of the CPA, the new mean-field
theory is endowed with a two-particle self-consistence. That is, the
two-particle irreducible vertices are determined from self-consistent
nonlinear equations.
 
The mean-field-like theory for the disorder driven vanishing of
diffusion of Ref.~\onlinecite{Janis04a} shows some unexpected
features. It contradicts the widely extended dogma that the weight of
the diffusion pole, i.~e., the number of diffusing particles, does not
depend on the disorder strength. The weight of the diffusion pole is
conserved if all the states near the Fermi energy are finite
combinations of Bloch waves for any configuration of the random
potential. Or equivalently if a Ward identity between self-energy and
the irreducible electron-hole vertex holds for all transfer
energies.\cite{Vollhardt80b,Janis02a} We demonstrated in
Ref.~\onlinecite{Janis04b} that once the electron-hole irreducible
vertex contains the so-called Cooper pole, the number of extended
states at the Fermi level decreases with increasing the disorder
strength. Hence, there is no chance for a theory with the Cooper pole
to fully satisfy the Ward identity between the averaged one- and
two-electron Green functions and to keep the number of diffusive
states independent of the disorder strength.

The aim of this paper is to clarify the ambiguities connected with the
mean-field concept applied to two-particle functions in disordered
electron systems. We delimit the content and the range of validity of
the two existing mean-field theories for the Anderson model of
noninteracting electrons --- the CPA and that of
Ref.~\onlinecite{Janis04a}. The common ground for both theories is the
limit to high spatial dimensions. We show that the ambiguity in the
identification of the mean-field theory for two-particle functions
results from two different ways how it can be derived: either via the
generating local functional and Ward identities, or via a direct
diagrammatic construction in high dimensions. The former construction,
CPA, is suitable only for one-electron spectral and thermodynamic
properties. Since the CPA lacks the electron-hole symmetry and does
not correctly reproduce the high-dimensional limit of two-particle
(vertex) functions, it becomes unreliable when applied to the
electrical conductivity and other transport properties. We demonstrate
that the latter approach, when properly formulated, can lead to a
theory being self-consistent and manifestly electron-hole symmetric at
the one- and as well at the two-particle level. The last two
conditions are necessary ingredients for a theory being able to
describe the Anderson localization transition.

The layout of the paper is as follows. In Section~\ref{sec:1P-self} we
summarize basic properties of the CPA defined from the limit to
infinite spatial dimensions. We show how the averaged grand potential
is derived from the local one-particle propagator and the self-energy.
The higher-order Green functions are then determined via local
external perturbations.  The mean-field theory with a two-particle
self-consistence is constructed in Section~\ref{sec:2P-self}. First,
inability of the CPA to reproduce the proper infinite-dimensional
limit for two-particle Green functions is demonstrated. Then, using
the parquet scheme and the electron-hole symmetry we derive a
self-consistent (nonlinear) equation for the irreducible two-particle
vertex. This equation is then solved at the mean-field level, i. e.,
in the leading nontrivial order of the high-dimensional limit. The
explicit form of the diffusion pole in high spatial dimensions with
its weight dependent on the disorder strength is finally obtained.
Consequences of our findings for understanding of the disorder-driven
vanishing of diffusion and Anderson localization are discussed in the
last Section~\ref{sec:Discussions}.

\section{Thermodynamic mean-field theory: One-particle self-consistence}
\label{sec:1P-self}

\subsection{One-particle functions and generating functional}
\label{sec:1p-functions}

To construct a comprehensive mean-field theory for thermodynamic
properties of a random system means to find an approximate
representation in closed form for the grand potential averaged over
random configurations
\begin{equation}\label{eq:Anderson-Av-FE}
\Omega(\mu)=-\frac{1}{\beta}\left\langle \ln \mbox{Tr}\exp \left\{ -\beta
\widehat{H} + \beta\mu\widehat{N}\right\} \right\rangle _{av}
\end{equation}
where $\mu$ is the chemical potential and $\widehat{N}$ is the
particle number operator. We will consider in this paper only a
noninteracting lattice electron gas scattered on random impurities and
described by the Anderson tight-binding Hamiltonian
\begin{eqnarray}\label{eq:AD_hamiltonian}
\widehat{H} &=&\sum_{<ij>}t_{ij}\widehat{c}_{i}^{\dagger}
\widehat{c}_{j}+\sum_iV_i \widehat{c}_{i }^{\dagger } \widehat{c}_{i}\ ,
\end{eqnarray}
where $V_i$ is a local, site-independent random potential.

It has become evident since the introduction of the concept of the
limit to infinite spatial dimensions in quantum itinerant systems that
a controllable comprehensive mean-field theory of itinerant models
should be defined via this formal limit.\cite{Metzner89} In high
spatial dimensions the diagonal (local) and off-diagonal (nonlocal)
elements of the one-particle propagator separate from each other. The
former are of order $O(d^0)$, while the latter vanish as $d^{-1/2}$,
where $d$ is the spatial dimension. The full one-particle propagator and
the self-energy have the following high-dimensional asymptotics
\begin{subequations}\label{eq:separation}
\begin{align}
  \label{eq:GF-separation}
  G&=G^{diag}[d^0]+G^{off}[d^{-1/2}]\ , \\ \label{eq:SE-separation}
  \Sigma &=\Sigma^{diag}[d^0] +\Sigma^{off}[d^{-3/2}]\ .
\end{align}
\end{subequations}

We can classify contributions to the many-body perturbation expansion
for the self-energy according to their high-dimensional asymptotic
contribution and obtain in the leading order a local approximation for
the irreducible part of the one-electron propagator. The interacting
part of the thermodynamic potential in infinite spatial dimensions is
then a functional of only $G^{diag}$ and
$\Sigma^{diag}$.\cite{Janis92}

In disordered systems the inter-particle interaction is replaced by
correlations between scatterings on impurities. The self-energy is
here a coherent potential of an effective homogeneous (nonrandom)
medium representing the effect of impurity scatterings on the motion
of electrons. Since the scatterings are static, we can find an
explicit representation of the averaged grand potential in infinite
spatial dimensions. We can write
\begin{subequations}\label{eq:MF}
 \begin{multline}\label{eq:MF-functional}
   \Omega_\mu\left[\widehat{G},\widehat{\Sigma}\right] =
   F\left\{\widehat{G}^{diag\ -1}+\widehat{\Sigma}^{diag}\right\}
   -\frac1{\beta}\,\mbox{Tr}\ln \widehat{G}^{diag}
   \\-\frac1{\beta}\,\mbox{Tr}\ln\left[\widehat{G}^{(0)-1}
     -\widehat{\Sigma}^{diag}+\mu\right]
 \end{multline}
 where we denoted
\begin{equation}\label{eq:MF-logarithm}
  F\left\{\widehat{X}\right\}= -\frac1{\beta}\left\langle\text{Tr}\ln\left[
  \widehat{X} - \widehat{V}\right]\right\rangle_{av}
\end{equation}
\end{subequations}
the local "interacting part" of the thermodynamic potential, in this
case the effect of multiple scatterings.\cite{Janis89} The trace
operator $\mbox{Tr}$ extends over the lattice space as well as over
the Matsubara frequencies. The only nonlocal contribution to the
generating functional $\Omega_\mu$ comes from the bare propagator
$\widehat{G}^{(0)}$. The site-diagonal (local) complex vectors
$G^{diag}_n(\mu)$ and $\Sigma^{diag}_n(\mu)$ in fermionic Matsubara
frequencies $(2n+1)\pi/\beta$ are variational parameters, the physical
values of which are attained at stationarity points of the generating
functional~\eqref{eq:MF}.

The defining equation for the local element of the averaged
one-particle propagator is obtained from an equation
$\delta\Omega_\mu/\delta\Sigma^{diag}_n(\mu) = 0$ and the self-energy
is determined from $\delta\Omega_\mu/\delta G^{diag}_n(\mu) = 0$.
After straightforward manipulations the two equations reduce to
\begin{subequations}\label{eq:SE-GF}
\begin{equation}\label{eq:SE-CPA}
  1=\left\langle\frac
1{1+\left[\Sigma_n(\mu)-V_i\right]G_n(\mu)}\right\rangle_{av}
\end{equation}
and
\begin{equation}\label{eq:1PGF-def}
G_n(\mu)=\frac1N\sum_{\mathbf{k}}G(\mathbf{k},i\omega_n) = \int \frac
{d\epsilon\rho(\epsilon)} {i\omega_n + \mu - \Sigma_n(\mu) - \epsilon}\ .
\end{equation}
\end{subequations}
We dropped the superscript $diag$ in the local functions and
introduced the electronic density of states $\rho(\epsilon)$. Due to
the static character of the impurity scatterings the stationarity
equations are diagonal in the Matsubara frequencies and can be solved
for each frequency independently. Inserting the solution for all
Matsubara frequencies to Eqs.~\eqref{eq:MF} we obtain the equilibrium
thermodynamic potential for noninteracting electrons scattered on
random impurities. The equilibrium thermodynamics of the systems is
then determined only by the local irreducible part of the averaged
one-particle resolvent. This irreducible part is self-consistently
determined from the Soven equation~\eqref{eq:SE-GF}.

\subsection{External sources and two-particle functions}
\label{sec:1P-WI}

Thermodynamics of disordered systems is not of much interest unless
inter-particle interactions are present. But even then the averaged
thermodynamic potentials depend on only one-electron functions.
One-electron functions, however, do not contain the complete
information about the behavior of statistical ensembles, in particular
of disordered systems. The equilibrium thermodynamic potentials do not
contain sufficient information from which we could derive transport
properties of the system and its response to weak external
electromagnetic perturbations.  To include the electrical conductivity
into the mean-field description, the thermodynamic construction from
the preceding subsection must be extended to include averaged
two-particle propagators.

Averaged two-particle propagators in disordered systems contain at
least two energy arguments (two in noninteracting and three in
interacting systems). The best way to guarantee that one- and
two-particle functions are approximated consistently within a single
approximate scheme is to use the Baym-Kadanoff concept of external
sources added to the equilibrium thermodynamic potential.\cite{Baym61}
To introduce higher-order Green functions with several energies
(chemical potentials) into the thermodynamic description we have to
replicate the original system so as for each energy we have an
independent replica of the original system, that is, of creation and
annihilation operators.

We replicate the creation and annihilation operators and introduce
external perturbations into the thermodynamic description via a
generalized grand potential of a $\nu$-times replicated system
$\Omega^{\nu}(E_1,E_2,\hdots,E_\nu;U)$ with $\nu$ chemical potentials
$E_1,\ldots,E_\nu$. An external perturbation $U$ is used to couple
different replicas and to break the initial replica independence. We
then can write
\begin{multline}
\label{eq:N-grand-potential} \Omega^{\nu}(E_1,E_2,\hdots,E_\nu;U)\\
=-\frac1{\beta}\left\langle\ln\mbox{Tr}\exp
  \left\{-\beta\sum_{i,j=1}^\nu\left(\widehat{H}^{(i)}_{AD}\delta_{ij}
    \right.\right.\right. \\ \left.\left.\left. -E_i
      \widehat{N}^{(i)}\delta_{ij} + \Delta\widehat{H}^{(ij)}
    \right)\right\} \right\rangle_{av}
\end{multline}
where we assigned to each replica characterized by energy (chemical
potential) $E_i$ a separate Hilbert space and denoted
$\Delta\widehat{H}^{(ij)}=\sum_{kl}U^{(ij)}_{kl}\widehat{c}_{k}^{(i)\dagger
}\widehat{c}^{(j)}_{l}$ an external perturbation to be set zero at the
end.  Thermodynamic potential $\Omega^{\nu}(E_1,E_2,\hdots,E_\nu;U)$
is a generating functional for averaged products of Green functions up
to the $\nu$th order. In practice, we will use linear-response theory
with one- and two-particle Green functions, i.~e.,
$\Omega^{\nu}(E_1,E_2,\hdots,E_\nu;U)$ is expanded up to $U^2$.
Therefore it is sufficient to introduce only two replicas.

In fact we are interested only in corrections to the products of the
averaged one-particle propagators expressed via vertex functions. The
two-particle vertex $\Gamma$ is defined from the two-particle resolvent
$G^{(2)}$ in momentum representation as
\begin{multline}\label{eq:2P_momentum}%
  G^{(2)}_{{\bf k}{\bf k}'}(z_1,z_2;{\bf q}) = G(\mathbf{k},z_1)
  G(\mathbf{k} + \mathbf{q},z_2) \left[\delta(\mathbf{k} -
    \mathbf{k}') \right. \\ \left.  + \Gamma_{{\bf k}{\bf
        k}'}(z_1,z_2;{\bf q}) G(\mathbf{k}',z_1) G(\mathbf{k}' +
    \mathbf{q},z_2)\right]\ .
\end{multline}

The external disturbance $U$ mixes different replicas and propagators
in the replicated space are matrices in the replica indexes. Since we
are interested only in the averaged two-particle functions, we can
resort to two energies and to a two-by-two matrix propagator
\begin{multline}
  \label{eq:1P-matrix}   \widehat{G}^{-1}({\bf k}_1,z_1,{\bf k}_2,z_2;U)
  = \widehat{G}^{(0)-1} + \widehat{U} - \widehat{\Sigma}\\
  = \begin{pmatrix}z_1-\epsilon({\bf k}_1) -\Sigma_{11}(U) & U -
    \Sigma_{12}(U)\\ U - \Sigma_{21}(U) & z_2-\epsilon({\bf k}_2)
    -\Sigma_{22}(U)
    \end{pmatrix}
\end{multline}
where $\epsilon({\bf k})$ is the lattice dispersion relation and the
self-energy elements $\Sigma_{ij}$ generally depend on both energies
$z_1,z_2$. The matrix $\widehat{G}$ represents the averaged resolvent
that is to be used in the grand potential $\Omega^{2}(E_1,E_2;U)$ from
Eq.~\eqref{eq:N-grand-potential}.  It is now a straightforward task to
derive a matrix Soven equation generalizing Eq.~\eqref{eq:SE-CPA}. We
obtain
\begin{multline}\label{eq:2CPA-equation}
  \widehat{G}(z_1,z_2;U)\\ =\left\langle
    \left[\widehat{G}^{-1}(z_1,z_2;U) +\widehat{\Sigma}(z_1,z_2;U)
      -\widehat{V}_i\right]^{-1}\right \rangle_{av}
\end{multline}
where $\widehat{G}(z_1,z_2;U)=N^{-2}\sum_{{\bf k}_1{\bf
    k}_2}\widehat{G}({\bf k}_1,z_1,{\bf k}_2,z_2;U)$ is the local
element of the matrix one-particle propagator. Inversions in
Eq.~\eqref{eq:2CPA-equation} have matrix character in the replica
space. The diagonal elements of the matrix
equation~\eqref{eq:2CPA-equation} determine the one-particle
propagators for energies $z_1$ and $z_2$. The off-diagonal elements,
proportional to the perturbation $U$, determine the local two-particle
resolvent that is defined as the coefficient at the linear term in the
expansion of the local matrix propagator $\widehat{G}(z_1,z_2;U)$ in
the external perturbation $U$. The local two-particle Green function
can be represented with the aid of the irreducible vertex
(two-particle self-energy) $\lambda$ via a Bethe-Salpeter equation.
We find from Eq.~\eqref{eq:2CPA-equation} that the Bethe-Salpeter
equation in the mean-field approximation reduces to an algebraic one
\begin{equation}\label{eq:local-vertex}
\gamma(z_1,z_2)=\frac{\lambda(z_1,z_2)}{1-\lambda(z_1,z_2)G(z_1)G(z_2)}
\ ,
\end{equation}
where $\gamma$ is the local part of the two-particle vertex $\Gamma$.
The irreducible vertex $\lambda$ in equilibrium ($U=0$) determined via
Eq.~\eqref{eq:local-vertex} obeys an equation
\begin{multline}\label{eq:2IP-vertex}
  \lambda(z_1,z_2)=\frac{\delta\Sigma_U(z_1,z_2)}{\delta
    G_U(z_1,z_2)}\bigg|_{U=0} =\frac 1{G(z_1)G(z_2)}\bigg[1- \\ \left.
    {\left\langle \frac 1{1+\left[\Sigma(z_1)-V_i\right]G(z_1)} \frac
        1{1+\left[\Sigma(z_2)- V_i\right]
          G(z_2)}\right\rangle^{-1}_{av}} \right] \ .
\end{multline}
We can easily verify that this equation coincides with the CPA
solution for the irreducible vertex
$\lambda(z_1,z_2)$.\cite{Velicky69}

There is no ambiguity in the mean-field construction of local one- and
two-particle functions. But a mean-field treatment has a physical
relevance only if it is able to produce nonlocal correlation
functions, the long-range fluctuations of which may significantly
influence the thermodynamic and dynamical behavior.  There is not,
however, a unique way how to generate the two-particle vertex with
non-local contributions within the local mean-field approach. The
simplest and most straightforward way is to use the Bethe-Salpeter
equation with the CPA irreducible vertex $\lambda$,
Eq.~\eqref{eq:local-vertex}, and to replace the local propagators with
the full nonlocal one-electron propagators $G(\mathbf{k},z)$. Such a
Bethe-Salpeter equation remains algebraic in momentum representation
and results in a two-particle vertex with only one transfer momentum.
We obtain
\begin{equation}\label{eq:CPA-vertex}
\Gamma^\pm(z_1,z_2;\mathbf{q}^\pm)=\frac{\lambda(z_1,z_2)}{1-
\lambda(z_1,z_2) \chi^\pm(z_1,z_2;\mathbf{q}^\pm)}
\end{equation}
where we denoted the two-particle bubble
\begin{equation}\label{eq:bubble}
  \chi^{\pm}(z_1,z_2;{\bf q})=\frac 1N\sum_{\bf k} G({\bf k},z_1)G({\bf
    q}\pm{\bf k},z_2)\ .
\end{equation}
The ambiguity in this definition of the full mean-field vertex is in
the type of nonlocal multiple scatterings we include into the
Bethe-Salpeter equation. They are here denoted by the superscript
$\pm$. The plus sign corresponds to multiple scatterings of
electron-hole pairs, while the minus sign to electron-electron pairs.
In case of elastic scatterings the electron-hole and electron-electron
bubbles produce numerically the same number. However, the difference
between the two types of pair scatterings lies in the respective
transfer momentum $\mathbf{q}^\pm$.  Using the notation for momenta in  the
two-particle resolvent from Eq.~\eqref{eq:2P_momentum} we have
$\mathbf{q}^+ =\mathbf{q}$ and $\mathbf{q}^- = \mathbf{q} + \mathbf{k}
+ \mathbf{k}'$.
 
This ambiguity in the definition of the mean-field two-particle vertex
is not usually acknowledged in the literature, since the electron-hole
scattering channel, relevant for the electrical conductivity, is
preferred and directly derived from the Baym-Kadanoff
approach.\cite{Velicky69} However, when the mean-field theory is
viewed upon as the limit to infinite spatial dimensions, both
electron-hole and electron-electron multiple scatterings possess the
same high-dimensional asymptotics.\cite{Janis99a} A priori, neither
electron-hole nor electron-electron multiple scatterings should be
discarded. The appropriate form of the vertex is then selected by the
physical quantities in which it appears, such as is the case of the
electrical conductivity.
 
Incapability of the thermodynamic mean-field theory to determine uniquely
the nonlocal part of the two-particle vertex results from the degeneracy
of the local theory with elastic scatterings only (noninteracting
systems). Multiple single-site scatterings, the only ones relevant in the
mean-field approach, are unable to distinguish between electrons and
holes. Only if we include explicitly scatterings on distinct lattice sites
we are able to distinguish between electrons and holes. Hence, the standard
thermodynamic mean-field theory of quantum itinerant systems uniquely
defines only the local two-particle vertex, while it remains ambiguous in
the determination of the full nonlocal two-particle vertex.

\section{Mean-field theory for vertex functions: Two-particle
  self-consistence} \label{sec:2P-self}

\subsection{Nonlocal contributions to the vertex function}
\label{sec:Nonlocal}
 
A rather inaccurate way to the momentum-dependent two-particle vertex
is not the only imperfection of the  thermodynamic mean-field
theory.  This theory completely fails to account for backscattering
effects, vertex corrections to the electrical conductivity, and the
Anderson metal-insulator transition. All these effects are induced by
strong nonlocal quantum coherence and spatial correlations reflected
in the momentum behavior of the two-particle (vertex) functions. To
capture these phenomena we have to go beyond a perturbative
description and to employ a self-consistent scheme for the
(irreducible) vertex functions. The best local approximation for the
irreducible vertex,  CPA, is non-self-consistent at the
two-particle level. We hence have to go beyond the CPA and include
nonlocal (long-range) contributions to the vertex function in a
non-perturbative manner. Thereby a question arises whether we are able
to reach a reasonably simple approximation with a two-particle
self-consistence that could be called a mean-field theory. It is clear
that such a theory must be momentum dependent, but the momentum
dependence should be reduced to a necessary minimum. We will
demonstrate in the next subsections that the desired momentum
dependence can be very effectively reduced by the asymptotic limit to
high spatial dimensions.

The best way to construct a mean-field-like approximation for
momentum-dependent functions is to build up the theory within a
self-consistent expansion around the CPA. If we denote the local CPA
one-particle propagator $G^{loc}(z)=N^{-1}\sum_{\bf k}G^{loc}({\bf
  k},z)$, the small parameter controlling the expansion around the CPA
is a perturbed propagator $\bar{G}({\bf k},z)=G({\bf
  k},z)-G^{loc}(z)$, where $G({\bf k},z)$ is the full one-electron
propagator. The CPA propagator $G^{loc}$ contains the local
self-energy $\Sigma^{loc}(z)$ from Eq.~\eqref{eq:SE-CPA}, while the
full one a self-energy $\Sigma({\bf k},z)$ that is to be determined
later from a Dyson equation.  We apply the expansion around the CPA to
two-particle functions, where the one-particle propagators will be
treated as external functions. It means, that we first disregard the
consistence between the one- and two-particle functions.  This
consistence will be restored later via Ward identities once a suitable
approximation for the vertex functions has been fixed.

We can classify {\em nonlocal} contributions to the two-particle
vertex by the type of the correlated two-particle propagation. We
either simultaneously propagate an electron and a hole or two
electrons (holes).  Diagrammatically it means that we connect {\em
  spatially distinct} two-particle scattering events with antiparallel
or parallel pairs of one-particle propagators. Multiple scatterings of
pairs of the same type define a channel of a two-particle
irreducibility. We call a diagram two-particle irreducible if it
cannot be split into separate parts by cutting simultaneously either
an electron-hole or an electron-electron pair of propagators. The two
definitions of the two-particle irreducibility lead to topologically
nonequivalent irreducible functions and to different Bethe-Salpeter
equations for the full vertex. In each Bethe-Salpeter equation the
two-particle functions are interconnected via one-particle propagators
in a different manner. We can generically represent the
channel-dependent Bethe-Salpeter equations as
\begin{multline} \label{eq:2P-reducible}
  \Gamma_{{\bf k}{\bf k}'}(z_+,z_-;{\bf
    q})=\bar{\Lambda}^\alpha_{{\bf k}{\bf k}'}(z_+,z_-;{\bf q})\\
  +\left[\bar{\Lambda}^\alpha \bar{G}\bar{G} \odot \Gamma
  \right]_{{\bf k}{\bf k}'}(z_+,z_-;{\bf q})\ .
\end{multline}
We used the symbol $\odot$ for the channel-dependent multiplication of
the two-particle functions represented by specific momentum
convolutions. Here $\bar{\Lambda}^\alpha$ is the irreducible vertex in
the $\alpha$-channel.

We will specify the momentum convolutions in the generic
Bethe-Salpeter equation~\eqref{eq:2P-reducible} for electron-hole and
electron-electron multiple scatterings only. There is also a third
two-particle channel, the so-called vertical channel with one-particle
self-correlating scatterings.\cite{Janis01b} These scatterings are,
however, unimportant for the phenomenon of Anderson localization,
since the corresponding two-particle propagator does not contain the
diffusion pole.

Using the notation from Eq.~\eqref{eq:2P_momentum} for the momentum
dependence of the two-particle functions we can write explicitly the
Bethe-Salpeter equation in the electron-hole channel with barred
functions as
\begin{subequations}\label{eq:BS}
\begin{multline}\label{eq:BS-eh}
  \Gamma_{\mathbf{k}\mathbf{k}'}(z_+,z_-;\mathbf{q}) =
  \bar{\Lambda}^{eh}_{\mathbf{k}\mathbf{k}'}(z_+,z_-;\mathbf{q})\\ +
  \frac 1N\sum_{\mathbf{k}''}
  \bar{\Lambda}^{eh}_{\mathbf{k}\mathbf{k}''}(z_+,z_-;\mathbf{q})
  \bar{G}_+(\mathbf{k}'') \bar{G}_-(\mathbf{k}'' + \mathbf{q})\\
  \times\Gamma_{\mathbf{k}''\mathbf{k}'}(z_+,z_-;\mathbf{q}) \ .
\end{multline}
The Bethe-Salpeter equation with the electron-electron multiple
scatterings then analogously reads
\begin{multline}\label{eq:BS-ee}
  \Gamma_{\mathbf{k}\mathbf{k}'}(z_+,z_-;\mathbf{q}) =
  \bar{\Lambda}^{ee}_{\mathbf{k}\mathbf{k}'}(z_+,z_-;\mathbf{q})\\ +
  \frac 1N\sum_{\mathbf{k}''}
  \bar{\Lambda}^{ee}_{\mathbf{k}\mathbf{k}''}(z_+,z_-;\mathbf{q} +
  \mathbf{k}' - \mathbf{k}'') \bar{G}_+(\mathbf{k}'')\\ \times
  \bar{G}_-(\mathbf{Q} - \mathbf{k}'')
  \Gamma_{\mathbf{k}''\mathbf{k}'}(z_+,z_-;\mathbf{q}+ \mathbf{k} -
  \mathbf{k}'')\ ,
\end{multline}
\end{subequations}
where we denoted $\mathbf{Q} = \mathbf{q} + \mathbf{k} + \mathbf{k}'$
the transfer momentum between the two electrons of the scattered
correlated pair. In these equations we abbreviated
$\bar{G}(\mathbf{k},z_\pm) \to \bar{G}_\pm(\mathbf{k})$.
 
Equations \eqref{eq:BS} constitute the fundamental building blocks for
the construction of systematic approximations for the two-particle
vertex.  They are analogues of the Dyson equation and enable us to
determine the full vertex from irreducible vertices. We hence can
apply the perturbation (diagrammatic) expansion to the two-particle
irreducible vertices, i.~e., two-particle self-energies.  As a first
step toward a mean-field-like theory for these vertices we have to
maximally simplify the momentum dependence of the vertex functions,
but still staying beyond the local CPA. This will be achieved by the
asymptotic limit to high (finite) spatial dimensions.

\subsection{Asymptotic limit to high lattice dimensions}
\label{sec:High-dim}

Bethe-Salpeter equations \eqref{eq:BS} use only off-diagonal
one-particle propagators and hence are suitable for performing the
limit to high spatial dimensions. We use the hypercubic lattice that
has a straightforward high-dimensional limit. The one-electron
propagator $\bar{G}$ has the following asymptotics
\begin{equation}\label{eq:1P-nonlocal}
\bar{G}(\mathbf{k},z) \doteq \frac{t}{\sqrt{d}}\sum_{\nu=1}^d \cos(k_\nu)
\int \frac{d\epsilon\rho(\epsilon)} {(z-\Sigma(z)-\epsilon)^{2}}\ ,
\end{equation}
where~$\Sigma(z)$ is the CPA ($d=\infty$)
self-energy.\cite{Mueller-Hartmann89}

The irreducible two-particle vertices must collapse to local
quantities in the limit to infinite dimensions. Since the
Bethe-Salpeter equations~\eqref{eq:BS} use only the off-diagonal
propagators vanishing in the limit $d=\infty$, both vertices
$\bar{\Lambda}^{eh}$ and $\bar{\Lambda}^{ee}$ must coincide with the
full local two-particle CPA vertex in $d=\infty$. We hence have
\begin{equation}\label{eq:LIR-high}
\bar{\Lambda}^{eh}_{\mathbf{k}\mathbf{k}'}(z_+,z_-;\mathbf{q}) =
\bar{\Lambda}^{ee}_{\mathbf{k}\mathbf{k}'}(z_+,z_-;\mathbf{q}) =
\gamma(z_+,z_-)\ .
\end{equation}
We further denote $\bar{\chi}(z_+,z_-;\mathbf{q}) =
\chi(z_+,z_-;\mathbf{q}) - G_+ G_-$ with $\chi(z_+,z_-;\mathbf{q}) =
\chi^+(z_+,z_-;\mathbf{q})$, $G_+ = G^{loc}(z_+)$, and $G_- =
G^{loc}(z_-)$. If we take into account only the electron-hole and
electron-electron multiple scatterings we can represent the leading
asymptotics of the full two-particle vertex in high dimensions as
follows
\begin{multline}\label{eq:vertex-high}
  \Gamma_{\mathbf{k}\mathbf{k}'}(z_+,z_-;\mathbf{q}) \doteq
  \gamma(z_+,z_-) + \lambda(z_+,z_-)\\ \times
  \left[\frac{\gamma(z_+,z_-) \bar{\chi}(z_+,z_-;\mathbf{q})} {1 -
      \lambda(z_+,z_-) \chi(z_+,z_-;\mathbf{q})}\right. \\ \left. +\
    \frac{\gamma(z_+,z_-)\bar{\chi}(z_+,z_-;\mathbf{Q})} {1 -
      \lambda(z_+,z_-) \chi(z_+,z_-;\mathbf{Q})}\right]\ .
\end{multline}
The standard nonlocal CPA vertex can be recovered from the above
expression if we neglect the contribution from the electron-electron
multiple scatterings, the second term in the brackets on the
right-hand side (r.h.s.)  of Eq.~\eqref{eq:vertex-high}. There is,
however, no reason for this neglect, since both terms within the
brackets on the r.h.s.  of Eq.~\eqref{eq:vertex-high} produce the same
asymptotic behavior in powers of the inverse dimension, $O(d^{-1})$.
Their only difference is in the momentum dependence.

The limit to infinite spatial dimensions reduces the momentum
dependence of two-particle functions but does not lead to a
mean-field-like approximation for vertices. There is no
self-consistence in the two-particle parameters and the local
irreducible vertices are determined from the CPA. We hence cannot
expect that this high-dimensional two-particle non-self-consistent
asymptotics would lead to major deviations from the CPA. The only
significant change in the vertex function, Eq.~\eqref{eq:vertex-high},
with respect to the CPA vertex from Eq.~\eqref{eq:CPA-vertex} are the
vertex corrections to the electrical conductivity in the form of
\textit{weak localization} properly described by multiple
electron-electron scatterings (maximally crossed diagrams).

\subsection{Parquet equations and electron-hole symmetry}
\label{sec:EH-parquet}

To go significantly beyond the CPA predictions for transport
properties and response functions we have to introduce a
self-consistence that would extend also to the two-particle
irreducible vertices. That is, the two-particle irreducible vertices
$\bar{\Lambda}^{eh}$ and $\bar{\Lambda}^{ee}$ should be determined
from nonlinear equations. This effect can be achieved by introducing
the so-called \textit{parquet equations}.  The concept of parquet
equations is based on the observation that two-particle reducible
diagrams in one scattering channel are irreducible in the other
\textit{distinguishable} scattering channels. Parquet equations were
introduced in many-body theories\cite{Dominicis62,Jackson82,Bickers91}
but recently they were adjusted also to disordered
systems.\cite{Janis01b} The reducible diagrams from one channel can
become irreducible in the other channels only if different channels
are indeed distinguishable or nonequivalent. The idea of parquet
equations cannot be applied to local propagators of noninteracting
particles with indistinguishable electrons and holes, hence within the
CPA. It, however, works very efficiently for nonlocal vertex functions
in the Anderson model of disordered electrons.

If we again take into account only the electron-hole and the
electron-electron scattering channels, we can write the basic parquet
equation for the full two-particle vertex in high dimensions
\begin{multline}\label{eq:Parquet-high}
  \Gamma_{\mathbf{k}\mathbf{k}'}(z_+,z_-;\mathbf{q})  =
  \bar{\Lambda}^{eh}_{\mathbf{k}\mathbf{k}'}(z_+,z_-;\mathbf{q}) \\+
  \bar{\Lambda}^{ee}_{\mathbf{k}\mathbf{k}'}(z_+,z_-;\mathbf{q}) -
  \gamma(z_+,z_-)\ .
\end{multline}
Equation \eqref{eq:Parquet-high} tells us that the full vertex is
decomposed into irreducible and reducible diagrams in the either
scattering channel and that the reducible diagrams consist of only the
irreducible diagrams from the other channel from which the completely
irreducible vertex, i. e., the vertex irreducible in both channels,
was subtracted. The limit to high lattice dimensions then determines
the completely irreducible vertex to be the full local CPA
($d=\infty$) vertex~$\gamma$.

Parquet equation \eqref{eq:Parquet-high} can now be used in the
Bethe-Salpeter equations \eqref{eq:BS} to exclude the full vertex
$\Gamma$ from them. Thereby we reach a closed set of nonlinear
integral equations for the irreducible vertices $\bar{\Lambda}^{eh}$
and $\bar{\Lambda}^{ee}$. This set of equations is generally not
solvable without further approximations. To approximate the resulting
parquet equations in a systematic and controlled way we again utilize
the mean-field idea --- limit to high dimensions. For two-particle
functions and parquet equations we have to use this limit only in the
asymptotic sense so that nonlocal fluctuations do not go lost
completely.

One can make an important observation in high spatial dimensions. The
off-diagonal one-particle propagators $\bar{G}$ behave in the leading
asymptotic order as Gaussian random variables with respect to momentum
summations.\cite{Janis04a} Using representation \eqref{eq:1P-nonlocal}
we can easily prove the following relations
\begin{subequations}\label{eq:Gauss-rules}
\begin{align}
  \frac 1N\sum_{\mathbf{q}'} \bar{\chi}(\mathbf{q}' + \mathbf{q})
  \bar{G}_\pm(\mathbf{q}' + \mathbf{k}) &\doteq
  \frac{Z}{4d}\,\bar{G}_\pm(\mathbf{q} - \mathbf{k})\ ,\\
  \frac 1N\sum_{\mathbf{q}}
  \bar{\chi}(\mathbf{q} + \mathbf{q}_1) \bar{\chi}(\mathbf{q} +
  \mathbf{q}_2) &\doteq \frac{Z}{4d}\,\bar{\chi}(\mathbf{q}_1 -
  \mathbf{q}_2)\ ,
\end{align}\end{subequations}
where we used abbreviations $Z=t^2\langle G_+^2\rangle\langle
G_-^2\rangle$ with $\langle G_\pm^2\rangle =
N^{-1}\sum_{\mathbf{k}}G_\pm(\mathbf{k})^2$.  The equalities in
Eq.~\eqref{eq:Gauss-rules} hold only within the leading asymptotic
order $d\to\infty$. The functions $\bar{G}_\pm$ and $\bar{\chi}$ form
a closed set of Gaussian random variables with respect to momentum
convolutions. We hence can use Eqs.~\eqref{eq:Gauss-rules} to simplify
the parquet equations for the irreducible vertices
$\bar{\Lambda}^{eh}$ and $\bar{\Lambda}^{ee}$.

Before we attempt to resolve the parquet equations in high dimensions, we
utilize the time-reversal invariance of the system. It is an important
feature of electron systems without spin- and orbital-dependent
scatterings. According to this invariance the physical (measurable)
results should not depend on the orientation of propagators. We hence can
write for one- and two-particle propagators
\begin{subequations}\label{eq:EH}
\begin{align}\label{eq:EH-GF}
  \bar{G}(\mathbf{k},z)& = \bar{G}(-\mathbf{k},z)\ , \\
  \Gamma_{\mathbf{k}\mathbf{k}'}(z_+,z_-;\mathbf{q})& =
  \Gamma_{\mathbf{k}\mathbf{k}'}(z_+,z_-;- \mathbf{Q})\nonumber\\
  & = \Gamma_{-\mathbf{k}'-\mathbf{k}}(z_+,z_-; \mathbf{Q})\ .
\end{align}
In case of the two-particle vertex, the time reversal was applied only
to one fermion propagator. The time reversal leaves the full
two-particle vertex invariant but it transforms the Bethe-Salpeter
equation~\eqref{eq:BS-eh} to Eq.~\eqref{eq:BS-ee} and vice versa.  It
means that the irreducible vertices transform as follows
\begin{align}\label{eq:EH-Lambda}
  \bar{\Lambda}^{ee}_{\mathbf{k}\mathbf{k}'}(z_+,z_-;\mathbf{q}) &=
  \bar{\Lambda}^{eh}_{\mathbf{k}\mathbf{k}'}(z_+,z_-;-\mathbf{Q})\nonumber\\
  &= \bar{\Lambda}^{eh}_{-\mathbf{k}'-\mathbf{k}}(z_+,z_-;\mathbf{Q})\
  .
\end{align}
\end{subequations}

The time-reversal (electron-hole) symmetry reduces the number of
parquet equations to just one nonlinear integral equation for a single
vertex that we define as
$\bar{\Lambda}_{\mathbf{k}\mathbf{k}'}(z_+,z_-;\mathbf{q}) \equiv
\bar{\Lambda}^{ee}_{\mathbf{k}\mathbf{k}'}(z_+,z_-;\mathbf{q}) =
\bar{\Lambda}^{eh}_{\mathbf{k}\mathbf{k}'}(z_+,z_-;-\mathbf{Q})$.  The
resulting equation for this vertex reads
\begin{multline}\label{eq:Parquet-single}
  \bar{\Lambda}_{\mathbf{k}\mathbf{k}'}(\mathbf{q}) = \gamma\\ +\frac
  1N\sum_{\mathbf{k}''}
  \bar{\Lambda}_{\mathbf{k}\mathbf{k}''}(-\mathbf{q} - \mathbf{k}
  -\mathbf{k}'')\bar{G}_+(\mathbf{k}'')\bar{G}_-(\mathbf{q}
  + \mathbf{k}'')\\
  \times\left[\bar{\Lambda}_{\mathbf{k}''\mathbf{k}'}(-\mathbf{q} -
    \mathbf{k}' -\mathbf{k}'') +
    \bar{\Lambda}_{\mathbf{k}''\mathbf{k}'}(\mathbf{q}) -
    \gamma\right]\ .
\end{multline}

Equation \eqref{eq:Parquet-single} can now be simplified in high
spatial dimensions by using the Gaussian summation rules,
Eqs.~\eqref{eq:Gauss-rules}. It is clear from these rules that the
fermionic momenta from different two-particle functions must be summed
independently in the leading asymptotic order. Any correlated momentum
summation involving two different two-particle functions costs a
factor $1/d$.  Then only the conserved bosonic transfer momenta
survive as in the case of the high-dimensional vertex from
Eq.~\eqref{eq:vertex-high}. We hence have to sum both sides of
Eq.~\eqref{eq:Parquet-single} over incoming and outgoing fermionic
momenta $\mathbf{k},\mathbf{k}'$ so as to extract the high-dimensional
limit of the irreducible vertex $\bar{\Lambda}$. To reach an equation
for the relevant variables we introduce
\begin{subequations}\label{eq:2PI-high}
\begin{equation}\label{eq:2PI-high-dim}
\bar{\Lambda}(\mathbf{q}) = \frac 1{N^2}\sum_{\mathbf{k}\mathbf{k}'}
\bar{\Lambda}_{\mathbf{k}\mathbf{k}'}(\mathbf{q}) \ .
\end{equation}
Further on we have in the leading order
\begin{equation}\label{eq:2PI-constant}
\frac 1{N^2}\sum_{\mathbf{k}\mathbf{k}'}
\bar{\Lambda}_{\mathbf{k}\mathbf{k}'}(\mathbf{q}+ \mathbf{k} + \mathbf{k}')
= \frac 1N \sum_{\mathbf{q}}\bar{\Lambda}(\mathbf{q})= \bar{\Lambda}_0\ .
\end{equation}\end{subequations}
Since the fermionic momenta from different two-particle functions are
summed independently in high spatial dimensions, the parquet equation
\eqref{eq:Parquet-single} reduces with the above definitions to
\begin{subequations}\label{eq:parquet-simplified}
\begin{equation}\label{eq:parquet-momentum}
\bar{\Lambda}(\mathbf{q}) = \gamma + \bar{\Lambda}_0 \frac {\bar{\Lambda}_0
\bar{\chi} (\mathbf{q})} {1 - \bar{\Lambda}_0\bar{\chi}(\mathbf{q})}\ .
\end{equation}
We see that the high-dimensional irreducible vertex is completely
determined from a single local (mean-field) parameter
$\bar{\Lambda}_0$ and the two-particle
bubble~$\bar{\chi}(\mathbf{q})$.  Summing both sides of
Eq.~\eqref{eq:parquet-momentum} over momenta we obtain an equation for
the local two-particle irreducible vertex
\begin{equation}\label{eq:parquet-local}
\bar{\Lambda}_0 = \gamma + \bar{\Lambda}^2_0 \frac 1N \sum_{\mathbf{q}}
\frac{ \bar{\chi}(\mathbf{q})}{1 - \bar{\Lambda}_0
\bar{\chi}(\mathbf{q})} \ .
\end{equation}\end{subequations}
Knowing the local part of the two-particle irreducible vertex
$\bar{\Lambda}_0$ we can reconstruct the full two-particle vertex in
high spatial dimensions. We have
\begin{multline}\label{eq:vertex-full}
  \Gamma_{\mathbf{k}\mathbf{k}'}(\mathbf{q}) = \gamma \\ + \Lambda_0
  \left[\frac{\bar{\Lambda}_0\bar{\chi}(\mathbf{q})} {1 - \Lambda_0
      \chi(\mathbf{q})} + \frac{\bar{\Lambda}_0\bar{\chi}(\mathbf{k} +
      \mathbf{k}' + \mathbf{q})} {1 - \Lambda_0 \chi(\mathbf{k} +
      \mathbf{k}' + \mathbf{q})}\right]
\end{multline}
where, in analogy to the non-self-consistent high dimensional vertex,
Eq.~\eqref{eq:vertex-high}, $\Lambda_0 = \bar{\Lambda}_0/(1 +
\bar{\Lambda}_0 G_+ G_-)$ and $\bar{\chi}(\mathbf{q}) =
\chi(\mathbf{q}) - G_+ G_-$.

Equations \eqref{eq:2PI-high} -- \eqref{eq:vertex-full} form an
approximation for two-particle functions with a single, local parameter
determined self-consistently. Such an approximation is a self-consistent
extension of the high-dimensional limit of the two-particle vertex,
Eq.~\eqref{eq:vertex-high}. We hence can call it a mean-field
approximation for two-particle functions of noninteracting disordered
electron systems. The self-consistently determined mean-field parameter
$\bar{\Lambda}_0$ exactly reproduces the leading $1/d$ correction to the
CPA irreducible vertex $\lambda$. The fundamental self-consistent equation
of this approximation, Eq.~\eqref{eq:parquet-local}, has a typical
mean-field character. That is, it can be used in any dimension. The
lattice structure enters the equation only through the momentum summation
running over the first Brillouin zone. Notice, however, that we cannot
reduce the momentum summation in two-particle functions to an integral
over the density of states. In high but finite spatial dimensions we have
to sum over momenta in Eq.~\eqref{eq:parquet-local} by using the
Gaussian rules only asymptotically  for $d\to\infty$.\cite{Janis04a}

\subsection{Self-energy and diffusion pole in high dimensions}
\label{sec:Diffusion-high}

The mean-field theory constructed in the preceding subsection is a
self-consistent approximation for the two-particle vertex where
one-particle propagators are assumed to be external functions. Such a
situation cannot be the final stage of the theory, since due to
conservation laws and thermodynamic consistence the one- and
two-particle functions are correlated. Actually, the electron-hole
irreducible vertex is connected with the self-energy for
noninteracting disordered electrons via the Vollhardt-W\"olfle Ward
identity\cite{Vollhardt80b}
\begin{multline}
  \label{eq:VWW-identity}   \Sigma^R({\bf k},E + \omega) - \Sigma^A({\bf
    k}, E) = \frac 1N \sum_{{\bf k}'}\Lambda^{RA}_{{\bf k}{\bf k}'}(E
  + \omega, E) \\ \times \left[G^R({\bf k}',E + \omega) - G^A({\bf
      k}', E) \right] \ .
\end{multline}
Here we denoted $\Sigma^{R},\Sigma^{A}$ the retarded and advanced
self-energy and $\Lambda^{RA}_{{\bf k}{\bf k}'}(E + \omega, E ) \equiv
\Lambda^{eh}_{{\bf k}{\bf k}'}(E + \omega + i0^+, E-
i0^+;\mathbf{0})$. Note that the irreducible vertex $\Lambda$ in
Eq.~\eqref{eq:VWW-identity} does not have bar, i.~e., it is defined
via the Bethe-Salpeter equation with the full one-electron propagators
$G^R,G^A$.  The Ward identity says that if we modify the electron-hole
irreducible vertex we have to change adequately the one-electron
self-energy and vice versa. We hence cannot approximate independently
the irreducible vertex without changing appropriately the self-energy.
If we have an analytic expression for the self-energy as a functional
of the one-electron resolvent, we can use a differential form of the
Ward identity \eqref{eq:2IP-vertex} to determine the irreducible
vertex. Then the vertex function is determined directly from the
one-electron propagator.  Such a construction of the irreducible
vertex does not lead to bifurcation points and multiple solutions,
i.~e., to a phase transition, unless we find them in the self-energy
itself. It is normally very difficult to determine bifurcation points
in the self-energy that is a bounded function. It is more convenient
to search for possible bifurcation points in two-particle functions
that may display divergences.

It is clear that we need self-consistent equations for functions that could
have multiple solutions. Such a self-consistent approximation for the
irreducible vertex was achieved in the preceding subsection. A
self-consistent equation for the irreducible vertex cannot be
derived from a self-energy directly. To achieve consistence between the
one- and two-particle functions in this case we have to determine the
self-energy as a functional of the irreducible vertex. This must be done
in concord with the Ward identity~\eqref{eq:VWW-identity}.

The self-energy as a functional of the irreducible vertex is
overdetermined from identity~\eqref{eq:VWW-identity}. We can
nevertheless use the Vollhardt-W\"olfle identity to determine the
self-energy from the vertex function as suggested in
Ref.~\onlinecite{Maleev75} and used in the parquet approach from
Ref.~\onlinecite{Janis01b}. We use only a specific element of the
Vollhardt-W\"olfle identity to determine the imaginary part of the
self-energy. In the mean-field case, where the irreducible vertex is
local, we can write a generalized CPA equation
\begin{subequations}\label{eq:KK}
\begin{align}\label{eq:KK-ImSE}
  \Im\Sigma^R(E) &= \Lambda^{RA}_0(E, E) \Im G^R(E)\ .
\end{align}
Consistence, or negative definiteness of the self-energy, demands that
the local element of the irreducible vertex $\Lambda^{RA}_0(E, E)=
N^{-2}\sum_{\mathbf{k}\mathbf{k}'}
\Lambda_{\mathbf{k}\mathbf{k}'}(E+i0^+,E-i0^+;\mathbf{0})$, determined
from the mean-field equation~\eqref{eq:parquet-local}, is positive.

We cannot find the real part of the self-energy directly from the
irreducible vertex. Instead, we use causality of the self-energy and
the Kramers-Kronig relation expressing the real part of an analytic
function as a Hilbert transform of its imaginary part. We have
\begin{align}\label{eq:KK-ReSE}
  \Re\Sigma(E) &= \Sigma_{\infty} + P\int_{-\infty}^\infty \frac
  {dE'}{\pi} \ \frac{\Im\Sigma^R(E')} {E' - E}\ .
\end{align}\end{subequations}
Equations \eqref{eq:KK} now determine the self-energy from the
mean-field (local) irreducible vertex $\Lambda_0$. Notice, however,
that to determine the self-energy at one frequency we have to know the
irreducible vertex in the whole frequency range. Equations~\eqref{eq:KK}
complete the mean-field theory for vertex functions,
Eqs.~\eqref{eq:2PI-high} and Eqs.~\eqref{eq:parquet-simplified}, and
make it a consistent approximation with proper analytic properties of
one- and two-particle Green functions.

The Vollhardt-W\"olfle identity, its specific form from
Eq.~\eqref{eq:KK-ImSE}, not only serves as a means for a consistent
determination of a causal self-energy from a given irreducible vertex,
but it is also indispensable for the existence of the diffusion pole
in the electron-hole correlation function. We now show in what form
the diffusion pole survives in the mean-field theory for vertex
functions with the self-energy determined by Eqs.~\eqref{eq:KK}.

The electron-hole correlation function is defined from the averaged
two-particle Green function as
\begin{equation} \label{eq:DRF-Phi} \Phi^{RA}_E(\mathbf{q},\omega) =
\frac 1{N^{2}}\sum_{\mathbf{k}\mathbf{k}'}
G^{(2)}_{\mathbf{k}\mathbf{k}'}(E+ \omega + i0^+ , E - i0^+;\mathbf{q}) \ .
\end{equation}
The two-particle Green function is evaluated with the full
two-particle vertex via Eq.~\eqref{eq:2P_momentum}.

From Eq.~\eqref{eq:KK-ImSE} and from $\chi^{RA}(\mathbf{0}) = \Im
G^R/\Im\Sigma^R$ we obtain that both the denominators in the
high-dimensional limit of the two-particle vertex $\Gamma^{RA}$ vanish
at zero transfer momenta. Hence, the mean-field approximation for
vertex functions contains the diffusion pole in the electron-hole
channel (first term in the brackets on the r.h.s. of
Eq.~\eqref{eq:vertex-full}) and the Cooper pole in the
electron-electron channel (second term). Only the diffusion pole
survives as a singularity in the electron-hole correlation function,
Eq.~\eqref{eq:DRF-Phi}.

To derive the low-energy behavior of the electron-hole correlation
function we denote
\begin{subequations}\label{eq:Diffusion-renorm}
\begin{equation}\label{eq:Diffusion-frequency}
A_E =  1 + 2 i\Im G^R(E)\ \frac{\partial
\Lambda^{RA}_0(E+\omega, E)}{\partial \omega}\bigg|_{\omega=0}
\end{equation}
and
\begin{equation}\label{eq:Diffusion-momentum}
D_E^0(\omega) = 2 \Im\Sigma^R(E) \Lambda^{RA}_0(E+\omega, E)
\frac{\partial \chi^{RA}(\mathbf{q})}{\partial( q^2)}\bigg|_{q=0}\ .
\end{equation}\end{subequations}
With these two definitions we find the high-dimensional asymptotics of
the low-energy limit of the electron-hole correlation function at zero
temperature to be
\begin{equation}\label{eq:Phi-high-dim}
\Phi^{RA}_E(\mathbf{q},\omega)\approx  \frac {2\pi n_E} {-i A_E\omega
+ D_E^0(\omega)\mathbf{q}^2}
\end{equation}
where $n_E$ is the density of states at the Fermi energy $E$.
 
The low-energy asymptotics of the electron-hole correlation function
serves as an important tool for testing consistence of approximations.
We find from gauge invariance and the (unrestricted) Ward identity
\eqref{eq:VWW-identity} that the electron-hole correlation function
should display the diffusion pole in form of
Eq.~\eqref{eq:Phi-high-dim} with $A_E=1$ for arbitrary disorder
strength. It then means that the low-energy behavior of the
electron-hole correlation function is controlled by a single
parameter, the bare dynamical diffusion constant
$D_E^0(\omega)$.\cite{Vollhardt80b,Janis02a} However, we already found
in Ref.~\onlinecite{Janis04a} that the constant $A_E$ increases with
the disorder strength and becomes infinite at the Anderson
localization transition. The disorder dependent weight of the
diffusion pole $n_E/A_E$ is the central unexpected feature of the
mean-field theory for vertex functions. This mean-field approximation
obeys the Vollhardt-W\"olfle identity only in the limit to zero
frequency, Eq.~\eqref{eq:KK-ImSE}, and not for finite energy
differences. We found a consistent explanation for the decrease of the
weight of the diffusion pole with increasing disorder
strength.\cite{Janis04b} The weight of the diffusion pole, $n_E/A_E$,
expresses a portion of extended (diffusive) states from all available
states at the Fermi energy determined by the density of states
calculated from the one-electron Green function, $n_E = -\Im
G^R(E)/\pi$ .

The dependence of the weight of the diffusion pole on the disorder
strength found in the mean-field theory for averaged two-particle
functions could be an artifact of this specific approximation. We
could still hope that the full exact solution recovers the invariant
weight of the diffusion pole expected from the unrestricted
conservation laws for averaged Green functions. Based only on
approximation-free arguments we found that the Vollhardt-W\"olfle
identity \eqref{eq:VWW-identity} for finite frequencies cannot be
fulfilled in any finite dimension if the electron-hole irreducible
vertex $\Lambda^{eh}$ contains the Cooper pole. Enforcing the full
form of the Vollhardt-W\"olfle identity with the Cooper pole in the
electron-hole irreducible vertex inevitably leads to a self-energy
being a nonanalytic function of its energy argument for almost all
Fermi energies within the energy bands.\cite{Janis03a,Janis04c} It
hence seems that the high-dimensional behavior of the two-particle
vertex and the disorder-dependent weight of diffusion pole, disclosed
by the asymptotic mean-field solution, are generic features of the
Anderson model of disordered electrons. At least for theories that can
be analytically continued from the limit to high spatial dimensions.

\section{Discussion and conclusions}
\label{sec:Discussions}

We presented in this paper two ways how to reach mean-field-like
approximations for noninteracting disordered electron systems. The
first one, being the standard thermodynamic mean-field theory known
from many-body systems, uses the limit to infinite spatial dimensions
applied to the generating, configurationally averaged thermodynamic
potential. The limit to infinite spatial dimensions enables one to
separate the diagonal and off-diagonal elements of the one-particle
propagator and its self-energy and to find an explicit representation
for the generating functional.  The local one-particle functions from
the generating functional serve as generalized variational parameters,
the physical values of which are determined from stationarity
equations for the generating functional. The higher-order Green and
vertex functions are determined from responses of the system to
\textit{local} external perturbations. In this way the construction of
a mean-field approximation is consistent and unambiguous in the
determination of one-particle functions as they are the only ones
entering the generating functional. The higher-order Green functions
are defined uniquely only in their local parts.

The nonlocal parts of two-particle Green functions are no longer
determined from the local thermodynamic theory uniquely. We can either
use the standard construction of Baym and Kadanoff or we can directly
apply the asymptotic limit to infinite spatial dimensions to
two-particle functions.  The outcome of these two constructions is not
identical. In the former way we miss some of the leading-order
(maximally crossed) diagrams and lose the electron-hole symmetry at
the two-particle level.  These deficiencies severely discredit
reliability of the thermodynamic mean-field theory in the calculation
of spatial coherence and transport properties of disordered systems.
To overcome these drawbacks we proposed another route toward a
mean-field-like approximation for vertex functions based on a direct
analysis of two-particle functions in high spatial dimensions.

The incapability of the thermodynamic mean-field theory to correctly
describe nonlocal correlations in two-particle functions is caused by
the degeneracy of local theories with elastic scatterings. Static
local approximations are unable to distinguish between electrons and
holes. Only quantum dynamics or multiple scatterings on spatially
distinct impurities can discern the motion of an electron from the
motion of a hole. The distinguishability of electrons and holes is of
principal importance for encountering backscattering effects and for a
two-particle self-consistence used in the construction of a mean-field
theory for vertex functions.

A mean-field approximation for vertex functions in disordered electron
systems was constructed from the\textit{ asymptotic} limit to high
spatial dimensions, where, unlike the thermodynamic mean field, the
lattice dimension is high but finite. Alike the strict limit
$d=\infty$, the asymptotic behavior in high dimensions leads to
significant simplifications in momentum convolutions that enable us to
reduce the approximation to a mean-field-like one with a local
generator determined from a self-consistent equation. We applied the
asymptotic limit directly within a diagrammatic expansion around the
CPA for two-particle functions.  The basic ingredients for the
construction were parquet equations for the two-particle irreducible
vertices from the electron-hole and the electron-electron scattering
channels. Using the electron-hole symmetry at the two-particle level
and the asymptotic limit to high dimensions we succeeded in producing
a self-consistent approximation for the local part of the
electron-hole irreducible vertex. It is a self-consistent $1/d$
extension of the CPA irreducible vertex.  The mean-field theory for
vertex functions then determines in a unique way the full two-particle
vertex that correctly reproduces the limit to infinite dimensions with
the electron-hole symmetry at both one- and two-particle levels.

The mean-field theory for vertex functions is an approximation for
two-particle functions. The one-electron functions, used as an input
for the two-particle equations, are then calculated from the vertex
function via a specific form of the Vollhardt-W\"olfle Ward identity
and the Kramers-Kronig relation. With this extension of the theory to
one-particle functions we accomplished an approximate description of
disordered systems fulfilling all consistency conditions on one- and
two-particle functions.

The most important achievement of this mean-field theory is its
ability to describe the disorder-induced vanishing of diffusion, that
is, the Anderson localization transition. This theory succeeded for
the first time to bridge qualitatively correctly the weak and the
strong disorder limits and to cover the split-band limit as well as
vanishing of diffusion. The other existing approaches have
concentrated on only one of the two phenomena. They either miss
the two-particle self-consistence or do not consistently match the
one- and two-particle functions.

The consistence between the one-electron self-energy and the
electron-hole irreducible vertex is essential for credibility of
approximate treatments of the Anderson metal-insulator transition.
Only with this relation correctly taken into account we obtain the
proper form of the diffusion pole and electron diffusion on long
distances.  In this respect the mean-field theory for vertex functions
leads, surprisingly against the common expectations, to a
nonconserving weight of the diffusion pole and its dependence on the
disorder strength. The thermodynamic mean-field theory and all other
approaches to the Anderson localization transition assume or use the
unrestricted form of the Ward identity \eqref{eq:VWW-identity} being a
consequence of conservation of the norm of the wave function for all
configurations of the random potential. We argued already earlier that
the Hilbert space of Bloch waves is incomplete in the sense that it
does not encompass the eigenstates of all configurations of the random
potential.  At a given energy we observe even in the metallic regime
macroscopically relevant numbers of configurations with localized
states.\cite{Janis04b}
 
Vanishing of the diffusion pole at the Anderson metal-insulator transition
and in the localized phase modifies the existing picture of the critical
behavior for vanishing of diffusion. In the standard approaches, such as
the nonlinear sigma model or the Vollhardt-W\"olfle self-consistent
approximation, the dynamical diffusion constant is the only parameter
controlling the low-energy behavior of the electron-hole correlation
function. In our mean-field theory  we have apart from the diffusion
constant also the weight of the diffusion pole that significantly
influences the description of the long-range correlations and diffusion.
In the critical region, however, only the constant $A_E\rightarrow\infty$
from Eq.~\eqref{eq:Diffusion-frequency} is relevant and all critical
scales can be derived from it. For instance the renormalized diffusion
constant $D_E = D_E^0/A_E$ vanishes at the localization transition with
the diverging parameter $A_E$, etc. It means that the Anderson
localization transition is compatible with a one-parametric scaling
theory.
 
Although the one-parameter scaling holds for the Anderson localization
transition, two relevant parameters in the critical region, $n_E/A_E$
and $D_E^0/A_E$, nevertheless lead to a modification of the critical
behavior deduced from the field-theoretic approaches. The two
parameters stand for two quantities influencing the electrical
conductivity. The former expresses the number of extended states at
the Fermi energy $E$ and the latter the averaged velocity of the
diffusive particles. Both quantities go simultaneously to zero at the
Anderson metal-insulator transition. It is straightforward to verify
that the mean-field theory for vertex functions predicts that $A_E\sim
|\lambda_c -\lambda|^{-1/2}$, where $\lambda$ is the bare disorder
strength and $\lambda_c$ its critical value. Having two vanishing
parameters we have to distinguish two types of the critical behavior.
First, we have properties of individual electrons.  One of them is
diffusion as seen from the semiclassical diffusion equation.  This is
quantitatively described by the renormalized diffusion constant $D_E
\sim |\lambda_c - \lambda|^{1/2}$. Second, we have statistical values
describing the disordered sample as a whole. Among them the averaged
conductivity is the most interesting one. It is proportional to the
product of the number of available diffusive states and the
renormalized diffusion constant, so that we have $\sigma \sim n_E
D_E^0/A^2 \sim |\lambda_c - \lambda|^1$.  Notice, however, that there
is not a direct relation between the diffusion constant and the
conductivity calculated from the Kubo formula, since due to deviations
from the Ward identity, the Einstein relation does not hold.  Finally
we can also deduce the critical exponent for the localization length
in the localized phase.  Its square is inversely proportional to the
order parameter, being the imaginary part of the local irreducible
vertex $\Im\Lambda^{RA}(E+ 0^+,E - 0^+) \sim |\lambda -
\lambda_c|^{1/2}$.  We hence obtain $\xi\sim |\lambda -
\lambda_c|^{-1/4}$.  The critical exponent for the conductivity equals
the mean-field exponent from the Vollhardt-W\"olfle theory but the
critical exponent for the localization length is one half of their
value.  We remind that the critical behavior obtained from the
mean-field theory for vertex functions holds only for dimensions
$d>d_u=4$.  We expect corrections to the mean-field critical behavior
in lower dimensions.
 
To conclude, we demonstrated that to reach a reliable quantitative
description of the Anderson localization transition one has to sum up
self-consistently nonlocal contributions to the electron-hole and the
electron-electron irreducible vertex functions. It can be achieved in
a mean-field manner via the asymptotic limit to high spatial
dimensions leading to a self-consistent extension of the CPA local
vertex. The most important conclusion of this mean-field theory of
Anderson localization is that the weight of the diffusion pole is not
conserved and that the diffusion pole is absent in the localized
phase. The decreasing weight of the diffusion pole with the increasing
disorder strength is a consequence of incompleteness of the Hilbert
space of Bloch waves. At any Fermi energy there are macroscopically
relevant numbers of configurations with localized as well as with
delocalized states. The number of configurations with extended states
decreases with increasing the disorder strength and vanishes at the
localization transition. This feature can essentially be checked by
other, e. g. numerical, means. Due to the disorder-dependent weight of
the diffusion pole the Einstein relation does not hold and we have to
distinguish individual and statistical transport properties of the
disordered systems. One should have this in mind when calculating the
critical behavior of the Anderson localization transition.

\section*{Acknowledgments}

Research on this problem was carried out within a project
AVOZ1-010-914 of the Academy of Sciences of the Czech Republic and
supported in part by Grant No. 202/04/1055 of the Grant Agency of the
Czech Republic.

\end{document}